# Post-Route Refinement for High-Frequency PCBs Considering Meander Segment Alleviation

Tsun-Ming Tseng
tsun-ming.tseng@tum.de

Bing Li
b.li@tum.de

*Tsung-Yi Ho
tyho@csie.ncku.edu.tw

Ulf Schlichtmann
ulf.schlichtmann@tum.de

Technische Universitaet Muenchen
Arcisstrasse 21, 80333 Munich, Germany

*National Cheng Kung University
No. 1, University Rd., Tainan City, Taiwan

## ABSTRACT
In this paper, we propose a post-processing framework which iteratively refines the routing results from an existing PCB router by removing dense meander segments. By swapping and detouring dense meander segments the proposed method can effectively alleviate accumulating crosstalk noise, while respecting pre-defined area constraints. Experimental results show more than 85% reduction of the meander segments and hence the noise cost.

## Categories and Subject Descriptors
J.6 [**Computer-aided Engineering**]: Computer-aided design (CAD)

## 1 Introduction
In high-frequency printed circuit boards, delay matching between bus signals has become a mainstream problem in modern PCB routers [1, 2, 5, 7]. In these papers, meander lines are commonly used to extend wire lengths to satisfy wire delay constraints, based on the concept that equal wire length should match the delays between different wires. Figure 1 illustrates a wire containing eight meander segments. However, it is shown in [3, 4, 6] that a signal across meander segments can have a speedup effect under high frequency circumstances, caused by the crosstalk noise of the same wire. In this paper, we propose a post-route refinement method to alleviate these dense meander segments in a given PCB design using swap, detour and shift techniques.

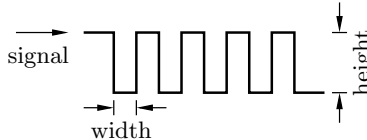

Figure 1: A wire with eight dense meander segments.

## 2 Proposed Method
In the following, we call the meander segments with small width *dense meander segments (dms)*, or *dense segments* if no ambiguity will ensue. The goal of the the proposed method is to reduce the number of the dense meander segments in a given PCB routing without changing the wire lengths or violating the given fixed outline. The concepts of the three techniques of the proposed method, swap, detour and shift, are explained in Section 2.1–2.3 and the complete flow applying these techniques is presented in Section 2.4.

### 2.1 Meander Segment Location Swap
In the first step, we swap the positions of the dense meander segments to lower their number on the wires. An example of this operation is shown in Figure 2, where the positions of the dense meander segments $dms2$ and $dms3$ are swapped. After this operation, $dms1$ and $dms4$ in Figure 2(a) are alleviated because the distances of their parallel wire segments are enlarged. During the swap step, no area overhead is required because we only take advantage of the area already used before.

### 2.2 Detour Generation
The detour technique increases wire lengths using free spaces so that dense meander segments elsewhere on the wires can be removed. An example of this technique is shown in Figure 3, where the dense meander segments are relaxed in (b) or (c) of Figure 3 by growing the required wire length with large distance between segments along the wires, called *detour*s in the following. In addition, it pushes multiple wires at the same time to avoid unnecessary area consumption. Consequently, *detour groups* are created on wires to replace the original dense meander segments, as shown in Figure 3.

In generating a detour, the simplest case of a pushable segment is a straight wire, which can be pushed upward or downward freely. However, the appearance of perpendicular wire segments, shown as dashed lines in Figure 4 and 5, may prevent the simple push operation, because the pushed segments may form new dense meander segments with the perpendicular wires. In Figure 4 the perpendicular segments are at the same direction of the straight horizontal wire. If the horizontal wire is short, one or more perpendicular segments are involved in this push operation, as shown in Figure 4(b)-(d). Note that the moving direction in Figure 4(d) can only be upward, because a downward push does not increase the wire length. In Figure 5 the perpendicular segments are at different sides of the straight horizontal wire. Similar to Figure 4(b) and (c), we can push the wires along one of the perpendicular segments. However, the push operation in Figure 4(d) can not be applied in this case, because moving the straight wire upward or downward can not compensate the shortened wire length, if some dense meander segments are removed elsewhere. In implementation, multiple wires are pushed at the same time to generate detour groups. Figure 6(a)-(d) exemplify this process using five wire segments, where wires are pushed downwards at the same and minimum wire distance between different wires are maintained. In Figure 6(d) the trim operation reduces the pushable region of the wire $s3$ from the left side to relax the occupied area. The retract operation in Figure 6(d) works similarly except that all the three wires, $s1$, $s3$ and $s5$ have their pushable regions reduced from the right side.

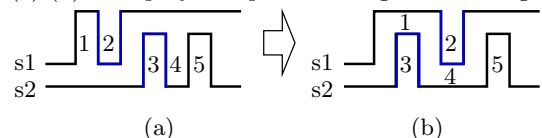

Figure 2: Meander segment swap.

Table 1: Results of routing refinement

| | initial routing | | | 100 iterations | | | | | 500 iterations | | | | |
|---|---|---|---|---|---|---|---|---|---|---|---|---|---|
| | $wires$ | $dms$ | $sum_H$ | $dms$ | $sum_H$ | $dms(\%)$ | $sum_H(\%)$ | $runtime(s)$ | $dms$ | $sum_H$ | $dms(\%)$ | $sum_H(\%)$ | $runtime(s)$ |
| case1 | 17 | 42 | 581 | 16 | 270 | 38.10 | 46.47 | 8.02 | 9 | 149 | 21.43 | 25.65 | 31.88 |
| case2 | 14 | 30 | 276 | 7 | 41 | 23.33 | 14.86 | 1.57 | 7 | 41 | 23.33 | 14.86 | 5.08 |
| case3 | 92 | 237 | 1789 | 117 | 869 | 49.37 | 48.57 | 18.56 | 88 | 566 | 37.13 | 31.64 | 82.53 |

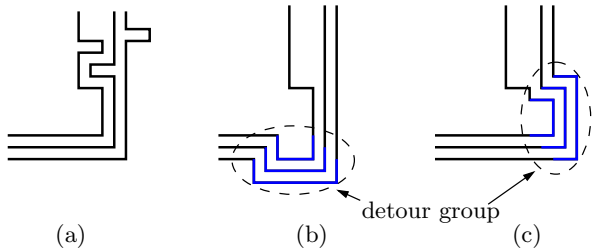

Figure 3: Example of detour generation.

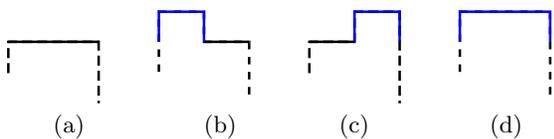

Figure 4: Pushable segment with perpendicular wires at the same side.

### 2.3 Segment Shifting

After the swap and detour operations, some wire segments may be moved to save routing area. For example, there is some space between the two leftmost wires in Figure 3(b) and (c). Therefore, the upper horizontal segments at the leftmost wire can be moved upward to free the space at the left side of this wire group, while keeping the length of the complete wire unchanged. After scanning the positions of wire segments, the shift operation contracts the movable wire segments in the area which may be needed by further detour groups to release more free space.

### 2.4 Complete Flow

In the implementation we first scan the given routing to identify existing dense segments using the coordinates of the bending points on the wires. For any pair of dense segments, we exchange their positions using the swap operation illustrated in Figure 2 if the overall speedup cost can be reduced. The detour and shift operations described in Section 2.2 and 2.3 are thereafter applied. The process described above runs many times until the given number of maximum iterations is reached or no improvement can be achieved in the current iteration.

## 3 Experimental Results

The experiments are performed using a computer with a 2.2 GHz CPU and 8 GB memory. The test cases are from [5]. The results of the three test cases after running the proposed method with 100 and 500 iterations are shown in Table 1. Beyond 500 iterations, no significant improvement was achieved further. In Table 1 the column $wires$ shows the number of wires and the columns $dms$ represent the numbers of dense meander segments in the routings. The sums of heights of these dense meander segments in the original routing and the routings improved by the proposed method are shown in the columns $sum_H$. $dms(\%)$ is the ratio of $dms$ after applying the proposed method over $dms$ in the original routing, and $sum_H(\%)$ is the ratio of the sums of heights. From these results, we can see that both the numbers of dense meander segments and the sums of their heights are reduced effectively.

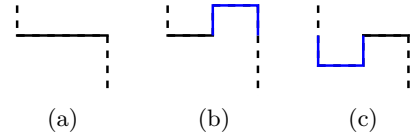

Figure 5: Pushable segment with perpendicular wires at different sides.

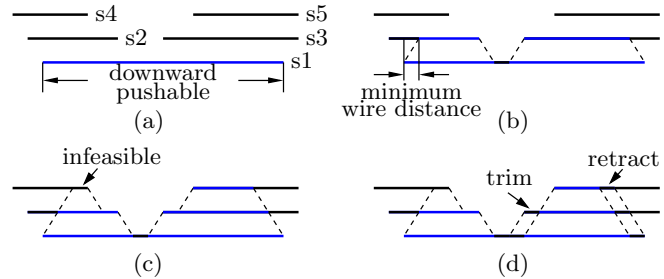

Figure 6: Finding a detour group.

## 4 Conclusion

In this paper we have explained a method to alleviate dense segments using techniques including swap, detour and shift. Experimental results confirm that the proposed method can effectively reduce the number of these dense meander segments and hence the speedup effect up to 85%.